\documentstyle[psfig]{mn}
\newif\ifAMStwofonts
\newcommand{\gapp}{\mbox{\raisebox{-0.3em}{$\stackrel{\textstyle >}{\sim}$}}}

\title[Environments of large radio galaxies]
      {A study of the environments of large radio galaxies using SDSS} 
\author[A. Pirya et al.]
       {A. Pirya,$^{1,2}$$\thanks{E-mail: akashpirya@gmail.com(AP), djs@ncra.tifr.res.in(DJS),
                                  msingh@aries.res.in(MS), chandolaharish@gmail.com(HCC)}$ 
        D. J. Saikia,$^{3,4}$ M. Singh$^{1}$ and H. C. Chandola$^{2}$ \\
$^{1}$ Aryabhatta Research Institute of Observational Sciences (ARIES),  Manora Peak, Nainital 263 129, India \\
$^{2}$ Department of Physics, Kumaun University, Nainital 263 002, India \\
$^{3}$ National Centre for Radio Astrophysics, TIFR, Pune University Campus, Post Bag 3, Pune 411 007, India \\
$^{4}$ Cotton College State University, Panbazar, Guwahati 781 001, Assam, India
}
\date{Accepted.    Received }

\pagerange{\pageref{firstpage}--\pageref{lastpage}}
\pubyear{  }

\begin{document}

\maketitle

\label{firstpage}

\begin{abstract}
The distributions of galaxies in the environments of 16 large radio sources 
have been examined using the Sloan Digital Sky Survey. In the giant radio
galaxy J1552+2005 (3C326) which has the highest arm-length ratio, the
shorter arm is found to interact with a group of galaxies which 
forms part of a filamentary structure. Although most large sources 
occur in regions of low galaxy density, the shorter arm is brighter in
most cases suggesting asymmetries in the intergalactic medium which may 
not be apparent in the distribution of galaxies. In two cases with strong
and variable cores, J0313+4120 and J1147+3501, the large flux density asymmetries 
are possibly also caused by the effects of relativistic motion.
\end{abstract}


%%%%%%%%%%%%%%%%%%%%%%%%%%%%%%%%%%%%%%%%%%%%%%%%%%%%%%%%%%%%%%%%%%%%%%%

\section{Introduction}
Large radio galaxies can be valuable probes of the intergalactic medium (IGM) 
and large-scale structures of the Universe. 
While the structures of the compact steep spectrum and giga-Hertz peaked
spectrum sources of subgalactic dimensions may be affected by interactions with
clouds in the interstellar medium of the host galaxies, radio
sources in clusters are affected by intracluster winds and peculiar velocities of
galaxies, giving rise to a wide variety of shapes such as the tailed radio sources
(e.g. Owen \& Rudnick 1976; Blanton et al. 2003; Giacintucci \& Venturi 2009; 
Mao et al. 2010). The structures of large radio sources, on the other hand, 
could be affected by the filaments, sheets and voids which make up the large-scale 
structure of our Universe.

Amongst large radio sources, the giant radio galaxies (GRGs) which are defined to be 
$\gapp$1 Mpc (H$_o$=71 km s$^{-1}$ Mpc$^{-1}$, $\Omega_m$=0.27, $\Omega_{vac}$=0.73, 
Spergel et al. 2003) are the largest single objects in the Universe.
The effect of galaxy distributions on the structures of individual GRGs
has been studied by a number of authors over the years, but these have been limited to
only a handful of sources. The sources studied include B0503$-$286 (Saripalli
et al. 1986), MSH 05$-$2{\it2} (Subrahmanyan et al. 2008), B0319$-$454 (Safouris et al. 2009), 
DA240 (Peng et al. 2004; Chen et al. 2011a),  NGC6251 (Chen et al. 2011b) 
and NGC315 (Chen et al. 2012). Chen et al. have studied the properties of the groups associated 
with NGC315 and NGC6251, and also highlighted a number of galaxies along the radio axis in 
DA240. Assuming that the galaxies trace the ambient IGM, Saripalli et al. (1986), 
Subrahmanyan et al. (2008) and Safouris et al. (2009) have explored and found evidence
of anisotropies in the medium affecting the radio structures.

In this paper we explore this theme for a sample of 16 large radio galaxies
which range from $\sim$700 to 4200 kpc using data from the Sloan Digital Sky Survey
(SDSS) Data Release 8 (DR8). 

%%%%%%%%%%%%%%%%%%%%%%%%%%%%%%%%%%%%%%%%%%%%%%%%%%%%%%%%%%%%%%%%%%%%%%%%%%%%%%%%%%%%%%

\section{Sample and galaxy counts}
For this study we have considered all large sources ($\gapp$700 kpc) listed by 
Ishwara-Chandra \& Saikia (1999), Lara et al. (2001), Machalski et al. (2001) and
Schoenmakers et al. (2000) which are within the area covered by SDSS DR8
and with the r-band magnitude of the host galaxy being $\leq$18. Since radio galaxies are 
usually associated with the brightest galaxy in a group or cluster, this should help us 
identify most of the associated members. For making a detailed multi-frequency study of 
these sources, we have also chosen them to have an integrated flux density at 1.4 GHz 
$\geq$100 mJy. We have excluded a couple of sources being studied independently from a 
sample compiled by Jamrozy et al. (in preparation).  This gives us a total of 16 sources (Table 1). 

%%%%%%%%%%%%%%%%%%%%%%%%%%%%%%%%%%%%%%%%%%%%%%%%%%%%%%%%%%%%%%%%%%%%%%%%%%%%%%%%%%%%%%%%%%%%%%%%%%%%%%%%%%55

\begin{table*}
\caption{The sample of the sources, where the column headings have their usual meaning. The total
flux densities at 1.4 GHz have been estimated from the NVSS images while the r-band magnitudes
(r-mag) of the host galaxies are from SDSS. }
\begin{tabular}{llllllllllll}
\hline
 Source & Alt. Name & RA(J2000.0) & DEC(J2000.0) & z & LAS & l & S$_{1.4}$ & r-mag & P$_{1.4}$ & Ref. \\
        &           & (hh mm ss)  & (dd mm ss)   &   & ($^{\prime\prime}$) & (kpc) & (mJy) & & (W/Hz) & \\
%(1)     & (2)       & (3)         & (4)          & (5)& (6)    & (7)       & (8)   & (9)   & (10)  & (11)  \\
\hline
 J0313+4120 & B0309+411   & 03:13:01.95 & +41:20:01.21 & 0.1340  & 532  & 1251 & 518   & 16.96 & 25.29 & S00 \\
 J0926+6519 &             & 09:26:00.82 & +65:19:23.00 & 0.1397  & 276  & 672  & 102   & 15.92 & 24.63 & L01 \\
 J1006+3454 & 3C236       & 10:06:01.73 & +34:54:10.52 & 0.0994  & 2302 & 4174 & 4516  & 15.09 & 25.97 & I99 \\
 J1113+4017 &             & 11:13:05.54 & +40:17:29.84 & 0.0745  & 637  & 891  & 240   & 15.01 & 24.45 & M01 \\
 J1147+3501 & 1144+352    & 11:47:22.13 & +35:01:07.54 & 0.0629  & 708  & 847  & 877   & 14.59 & 24.88 & I99 \\
 J1220+6341 &             & 12:20:36.41 & +63:41:44.48 & 0.1876  & 250  & 777  & 260   & 17.25 & 25.41 & L01 \\
 J1247+6723 & VII Zw 485  & 12:47:33.31 & +67:23:16.72 & 0.1072  & 634  & 1229 & 393   & 16.03 & 24.98 & L01 \\
 J1311+4058 & B1309+412   & 13:11:43.08 & +40:58:59.78 & 0.1103  & 368  & 731  & 596   & 16.28 & 25.18 & S00 \\
 J1328$-$0307 &           & 13:28:34.36 &$-$03:07:44.78& 0.0853  & 818  & 1293 & 218   & 16.70 & 24.51 & M01 \\
 J1342+3758 &             & 13:42:54.51 & +37:58:18.81 & 0.2270  & 657  & 2369 & 141   & 17.86 & 25.16 & M01 \\
 J1345+3952 & B1342+401   & 13:45:03.57 & +39:52:31.42 & 0.1612  & 307  & 843  & 178   & 17.24 & 24.96 & M01 \\
 J1400+3019 & B1358+305   & 14:00:43.43 & +30:19:18.59 & 0.2060  & 542  & 1813 & 449   & 17.38 & 25.57 & S00 \\
 J1428+2918 & B1426+295   & 14:28:19.23 & +29:18:44.19 & 0.0870  & 850  & 1368 & 437   & 15.33 & 24.83 & M01 \\
 J1453+3308 & 4C +33.33   & 14:53:02.86 & +33:08:42.41 & 0.2481  & 361  & 1391 & 459   & 17.74 & 25.74 & M01 \\
 J1552+2005 & 3C326       & 15:52:09.11 & +20:05:48.17 & 0.0898  & 964  & 1595 & 2347  & 15.88 & 25.59 & I99 \\
 J1635+3608 &             & 16:35:22.54 & +36:08:04.99 & 0.1650  & 385  & 1078 & 100   & 16.74 & 24.76 & M01 \\

\hline
\end{tabular}
References: S00: Schoenmakers et al. 2000; L01: Lara et al. 2001; I99: Ishwara-Chandra et al. 1999; 
            M01: Machalski et al. 2001  \\
\end{table*}

%%%%%%%%%%%%%%%%%%%%%%%%%%%%%%%%%%%%%%%%%%%%%%%%%%%%%%%%%%%%%%%%%%%%%%%%%%%%%%%%%%%%%%%%%%%%%%%%%%%%%%%%%%

%%%%%%%%%%%%%%%%%%%%%%%%%%%%%%%%%%%%%%%%%%%%%%%%%%%%%%%%%%%%%%%%%%%%%%%%%%%%%%%%%%%%%%%%%%%%%%%%

Using SDSS DR8 data of the Baryon Oscillation Spectroscopic Survey or BOSS (Aihara et al. 2011), 
we have initially identified all the galaxies within a box of $\sim$6 times the source size 
centred on the identified galaxy, and up to 5 mag fainter than the host galaxy. The box widths and the 
number of galaxies within them are listed in Table 2. Almost all these galaxies have a spectroscopic
redshift in SDSS DR8. We have used these to estimate the recession velocity, $cz$, and identify
all the galaxies which are within $\pm$1500 and $\pm$2500 km s$^{-1}$ of the host galaxy. Since
interacting clusters may sometimes have large velocity dispersion (e.g. Mao et al. 2010), we have 
considered values as large as $\pm$2500 km s$^{-1}$. The number of galaxies within these limits 
is also listed in Table 2. We have also estimated the separation and flux density ratios of 
the outer lobes for these sources (Table 3).

%%%%%%%%%%%%%%%%%%%%%%%%%%%%%%%%%%%%%%%%%%%%%%%%%%%%%%%%%%%%%%%%%%%%%%%%%%%%%%%%%%%%%%%%%%%%%%%%%%%55

\begin{table}
\caption{Summary of galaxy counts.
Column 1: Source name;
columns 2 and 3: the box width in arcsec and kpc respectively;  
column 4: the number of fainter galaxies within 5 mag of the host galaxy in the r band; 
columns 5 and 6: the number of galaxies within $\pm$1500 km s$^{-1}$ and $\pm$2500 km s$^{-1}$ of 
              the host galaxy, respectively.}

\begin{tabular}{lllllll}
\hline
  Source      & width                 & width   & N$^{r+5}_{gal}$   & N$^{\pm1500}_{gal}$    & N$^{\pm2500}_{gal}$     \\
%              & width               & width &                   &                        &                         \\ 
              & ($^{\prime\prime}$) & (kpc) &                   &                        &                         \\  
 (1)          & (2) & (3)   & (4)   & (5) & (6)    \\ 
\hline
 J0313+4120   & 3192  & 7504  &   3  &  2    & 2   \\   
 J0926+6519   & 1656  & 4034  &   9  &  5    & 5   \\   
 J1006+3454   & 13812 & 25041 & 1053 & 110   & 156 \\   
 J1113+4017   & 3822  & 5343  &  231 & 158   & 167 \\   
 J1147+3501   & 4248  & 5081  &  118 & 15    & 24  \\   
 J1220+6341   & 1500  & 4659  &  10  &  1    & 1   \\   
 J1247+6723   & 3804  & 7372  &  28  &  2    & 2   \\   
 J1311+4058   & 2208  & 4387  &  26  & 6     & 6   \\   
 J1328$-$0307 & 4908  & 7760  &  130 & 20    & 22  \\   
 J1342+3758   & 3942  & 14211 &  17  &  1    &  1  \\   
 J1345+3952   & 1842  & 5058  &  14  &  2    & 2   \\   
 J1400+3019   & 3252  & 10875 &  40  &  5    & 7   \\   
 J1428+2918   & 5100  & 8206  &  142 & 20    & 25  \\   
 J1453+3308   & 2166  & 8348  &  9   &  1    & 1   \\   
 J1552+2005   & 5784  & 9573  &  258 & 45    & 60  \\   
 J1635+3608   & 2310  & 6466  &  29  & 5     & 6   \\   
\hline
\end{tabular}

\end{table}

%%%%%%%%%%%%%%%%%%%%%%%%%%%%%%%%%%%%%%%%%%%%%%%%%%%%%%%%%%%%%%%%%%%%%%%%%%%%%%%%%%%%%%%%%%%%%%%%%%%%%%%%%%%%%%%%%

%%%%%%%%%%%%%%%%%%%%%%%%%%%%%%%%%%%%%%%%%%%%%%%%%%%%%%%%%%%%%%%%%%%%%%%%%%%%%%%
\begin{figure*}
\vbox{
%\vskip -0.5cm
\hbox{
  \psfig{file=1552_2005_CORRECTED_3RAD_5R_2500_1500.PS,width=3.35in,angle=0}
  \psfig{file=1113_4017_3RAD_5R_2500_1500.PS,width=3.33in,angle=0}
     }
\vskip -0.2cm
\hbox{
\hspace{0.4cm}
  \psfig{file=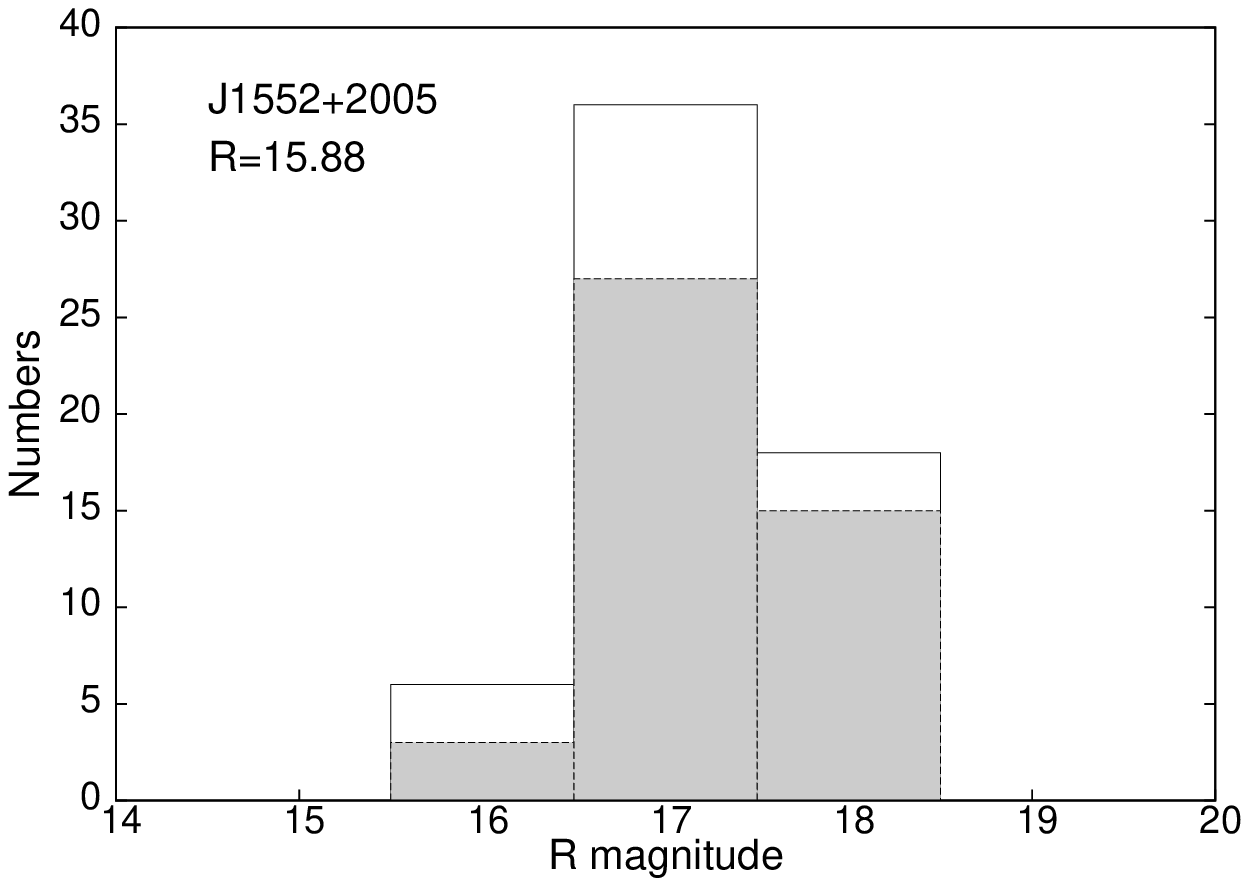,width=1.56in,angle=0}
  \psfig{file=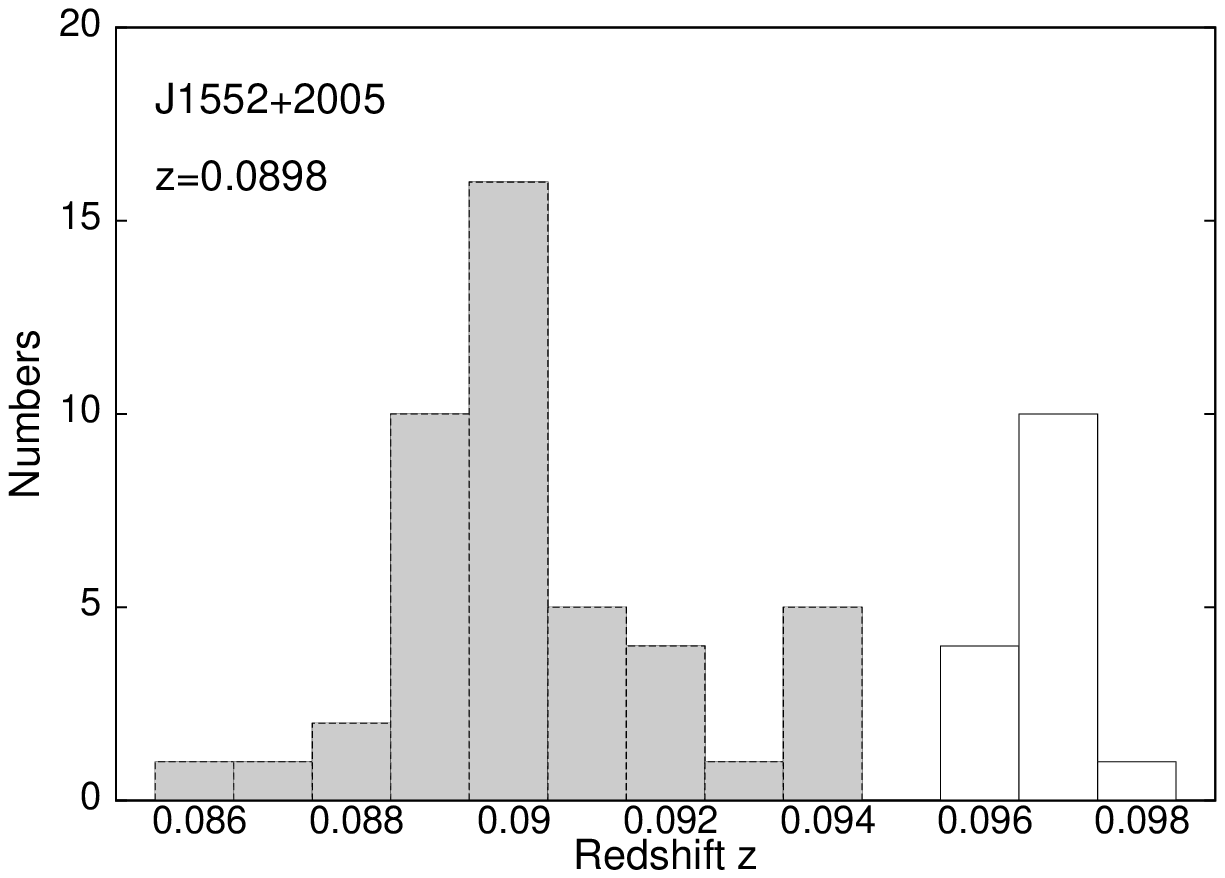,width=1.56in,angle=0}
\hspace{0.3cm}
  \psfig{file=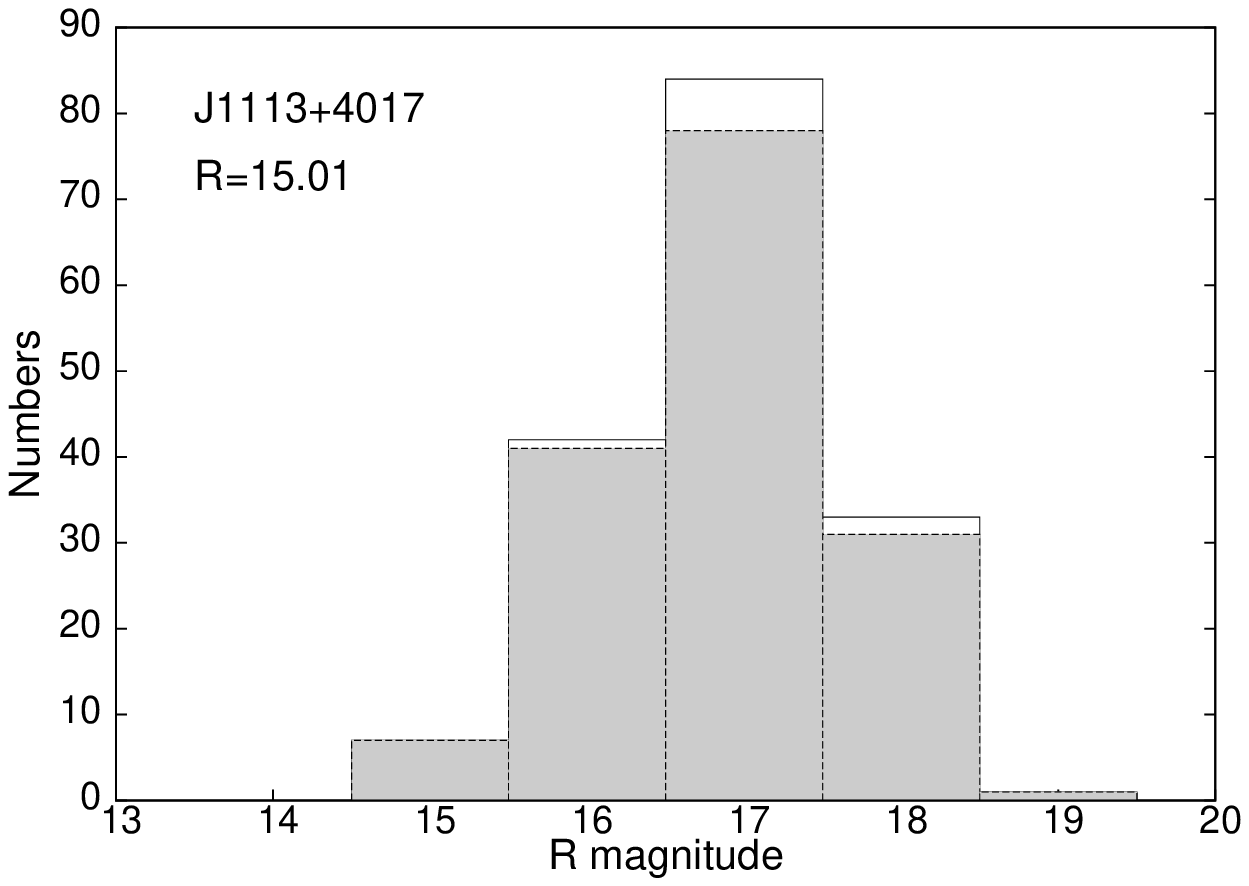,width=1.56in,angle=0}
  \psfig{file=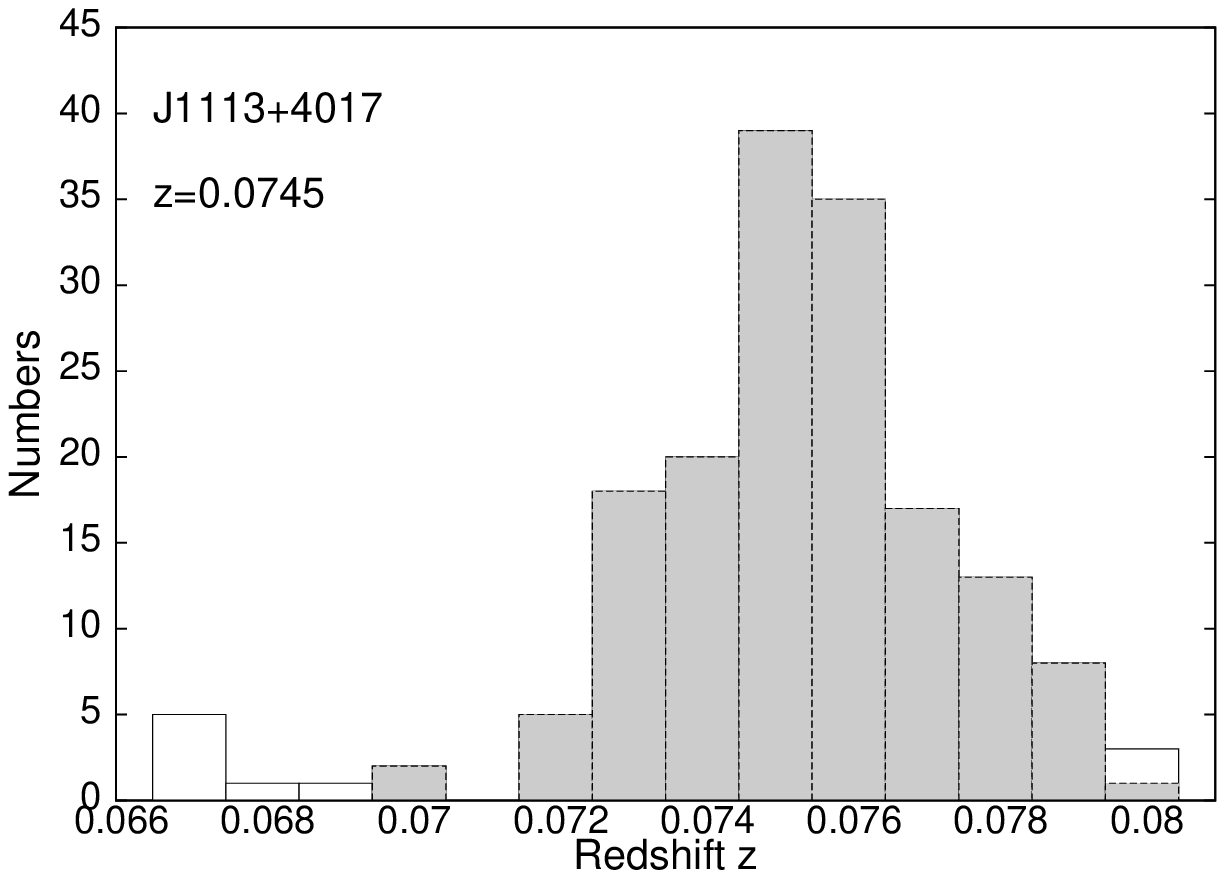,width=1.56in,angle=0}
     }
      }
\caption[]{Radio images towards J1552$+$2005 and J1113$+$4017.
           The image towards J1552$+$2005 is from the VLA Low-frequency Sky Survey (VLSS; Cohen et al. 2007),
           while the one towards J1113$+$4017 is from the NRAO VLA Sky Survey (NVSS; Condon et al. 1998). 
           In all the Figures ths symbols are as follows; $+$: the positions of the host galaxies;
           $\times$: the positions of galaxies within $\pm$1500 km s$^{-1}$ of the host 
           galaxies; $\Box$: positions of galaxies whose absolute 
           recession velocities lie between 1500 and 2500 km s$^{-1}$ of the host galaxies.
           The magnitude and redshift distributions are given below each contour map.
           In all the figures, the shaded boxes show the distributions for sources within
           $\pm$1500 km s$^{-1}$ of the host galaxies while the unshaded boxes show the distributions
           for sources whose absolute relative velocities lie between 1500 and 2500 km s$^{-1}$.   
           }
\end{figure*}

%%%%%%%%%%%%%%%%%%%%%%%%%%%%%%%%%%%%%%%%%%%%%%%%%%%%%%%%%%%%%%%%%%%%%%%%%%%%%%%%%%%%%%%%%%%%%%%%%%%5

\begin{table}
\caption{Symmetry parameters of the sources.
Column 1: Source name;
column 2: R$_{\theta}$, the arm-length ratio defined to be $>$1;
column 3: R$_{s}^{lobe}$, ratio of the flux density of the longer lobe to that of 
          the shorter one at 1.4 GHz;
column 4: the longer lobe;
column 5: the fraction of core emission at 1.4 GHz;
column 6: references.
   }

\begin{tabular}{lllllll}
\hline
 Source Name & R$_{\theta}$ & R$_{s}^{lobe}$  & lobe  & f$_{c}$ & Ref.  \\
(1)          & (2)          & (3)               & (4)   &  (5)    & (6)   \\
\hline                                                 
J0313+4120   & 1.67         & 0.16              & SW    & 0.709   & dB89, S00  \\
J0926+6519   & 1.01         & 0.91              & SW    & 0.090   & NVSS  \\
J1006+3454   & 1.61         & 0.58              & SE    & 0.725   & S80   \\
J1113+4017   & 1.16         & 0.69              & SW    & 0.045   & NVSS  \\
J1147+3501   & 1.20         & 2.09              &  E    & 0.698   & S99   \\
J1220+6341   & 1.40         & 0.96              & SE    & 0.012   & NVSS  \\
J1247+6723   & 1.07         & 0.90              & SE    & 0.654   & NVSS  \\
J1311+4058   & 1.01         & 0.73              &  S    & 0.001   & V89, NVSS  \\
J1328$-$0307 & 1.15         & 0.50              & NE    & 0.040   & M99, NVSS  \\
J1342+3758   & 1.38         & 0.69              & NE    & 0.009   & J05  \\
J1345+3952   & 1.16         & 0.83              &  E    & 0.027   & NVSS  \\
J1400+3019   & 2.47         & 0.46              &  S    & 0.004   & P96   \\
J1428+2918   & 1.31         & 0.96              & SW    & 0.028   & S00, NVSS  \\
J1453+3308   & 1.31         & 0.59              &  S    & 0.009   & K06   \\
J1552+2005   & 3.10         & 0.53              &  W    & 0.002   & M97   \\
J1635+3608   & 2.04         & 0.74              & SW    & 0.039   & NVSS  \\ 
\hline
\end{tabular}

References: dB89: de Bruyn 1989; S00: Schoenmakers et al. 2000; NVSS: NVSS survey; 
            S80: Strom \& Willis 1980; S99: Schoenmakers et al. 1999; V89: Vigotti et al. 1989;
            M99: Machalski \& Condon 1999; J05: Jamrozy et al. 2005; P96: Parma et al. 1996; 
            K06: Konar et al. 2006; M97: Mack et al. 1997
            \\

\end{table}

%%%%%%%%%%%%%%%%%%%%%%%%%%%%%%%%%%%%%%%%%%%%%%%%%%%%%%%%%%%%%%%%%%%%%%%%%%%%%%%%%%%%%%%%%%%%

%%%%%%%%%%%%%%%%%%%%%%%%%%%%%%%%%%%%%%%%%%%%%%%%%%%%%%%%%%%%%%%%%%%%%%%%%%%%%%%
\begin{figure*}
\vbox{
%\vskip -0.5cm
\hbox{
  \psfig{file=0926_6519_3RAD_5R_2500_1500.PS,width=3.51in,angle=-90}
  \psfig{file=1006_3454_CORRECTED_3RAD_5R_2500_1500.PS,width=3.21in,angle=0}
     }
\vskip -0.3cm
\hbox{
\hspace{0.4cm}
  \psfig{file=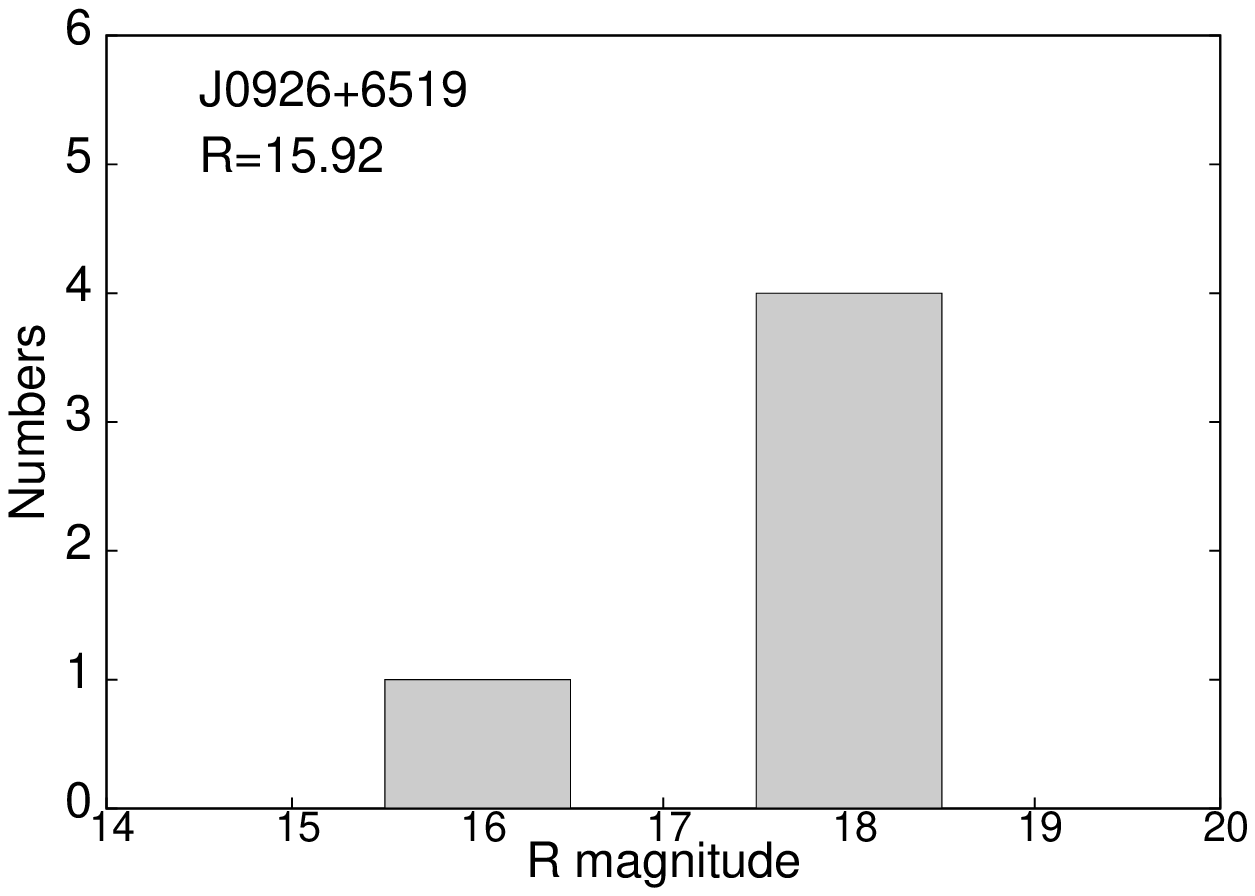,width=1.6in,angle=0}
  \psfig{file=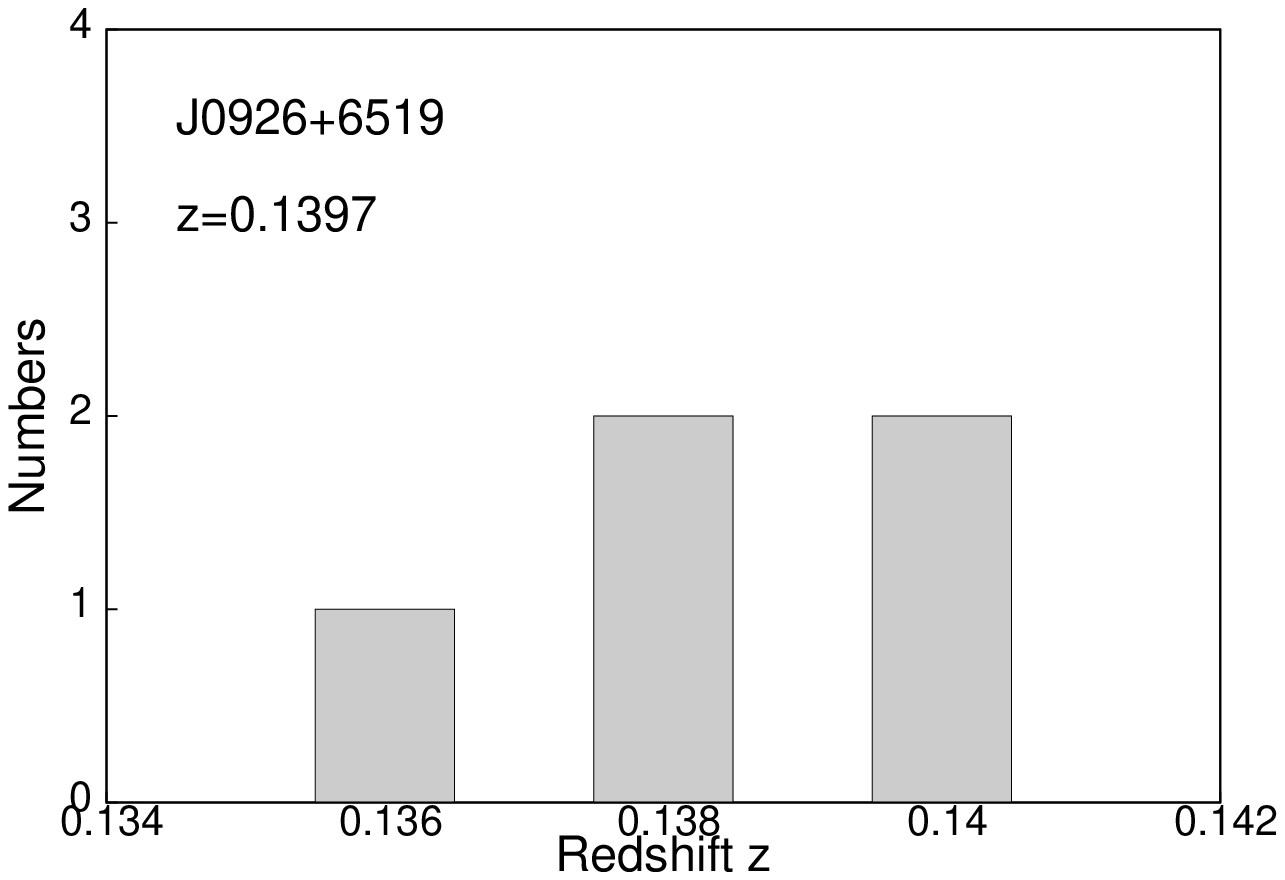,width=1.6in,angle=0}
\hspace{0.3cm}
  \psfig{file=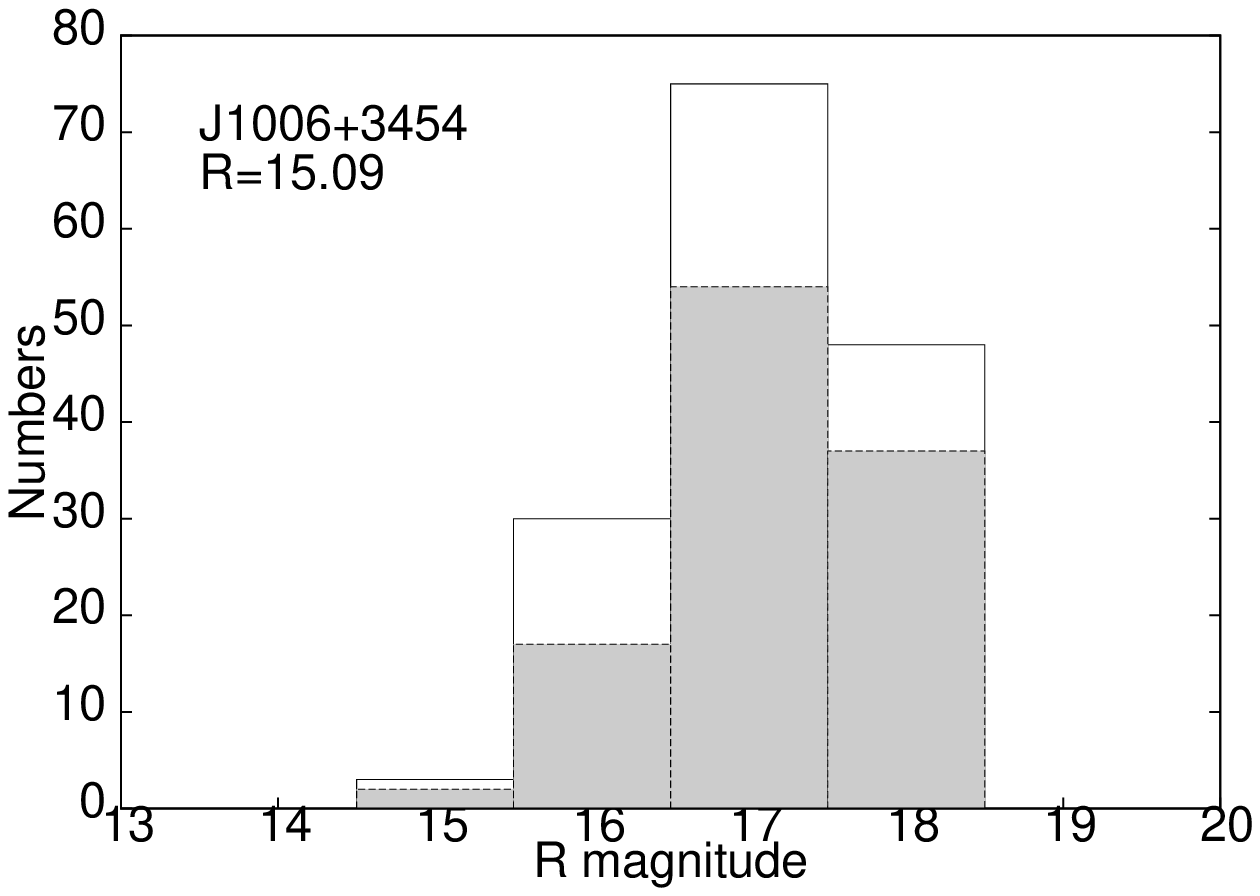,width=1.6in,angle=0}
  \psfig{file=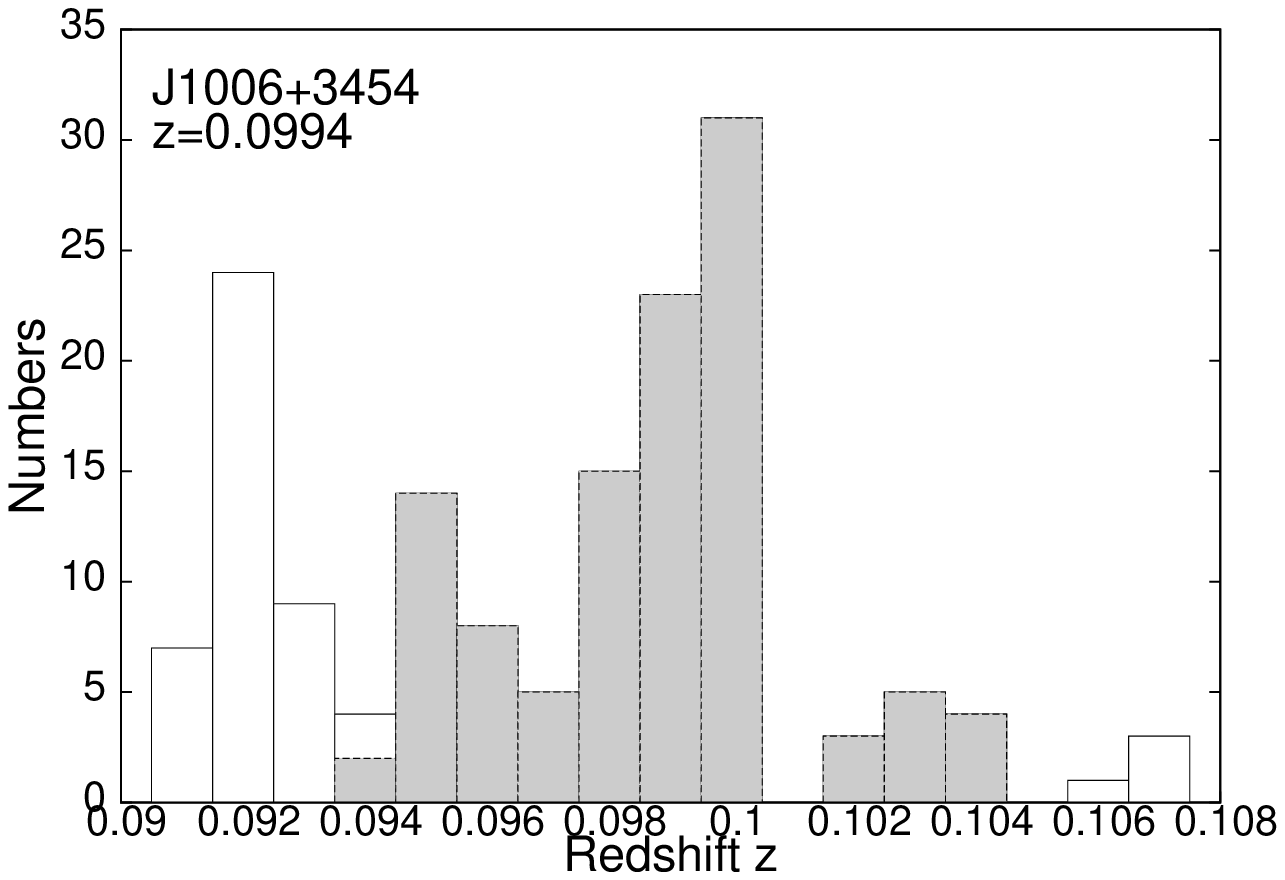,width=1.6in,angle=0}
     }
\hbox{
  \psfig{file=1147_3501_CORRECTED_3RAD_5R_2500_1500.PS,width=3.55in,angle=-90}
  \psfig{file=1311_4058_3RAD_5R_2500_1500.PS,width=3.2in,angle=0}
     }
\vskip -0.3cm
\hbox{
\hspace{0.5cm}
  \psfig{file=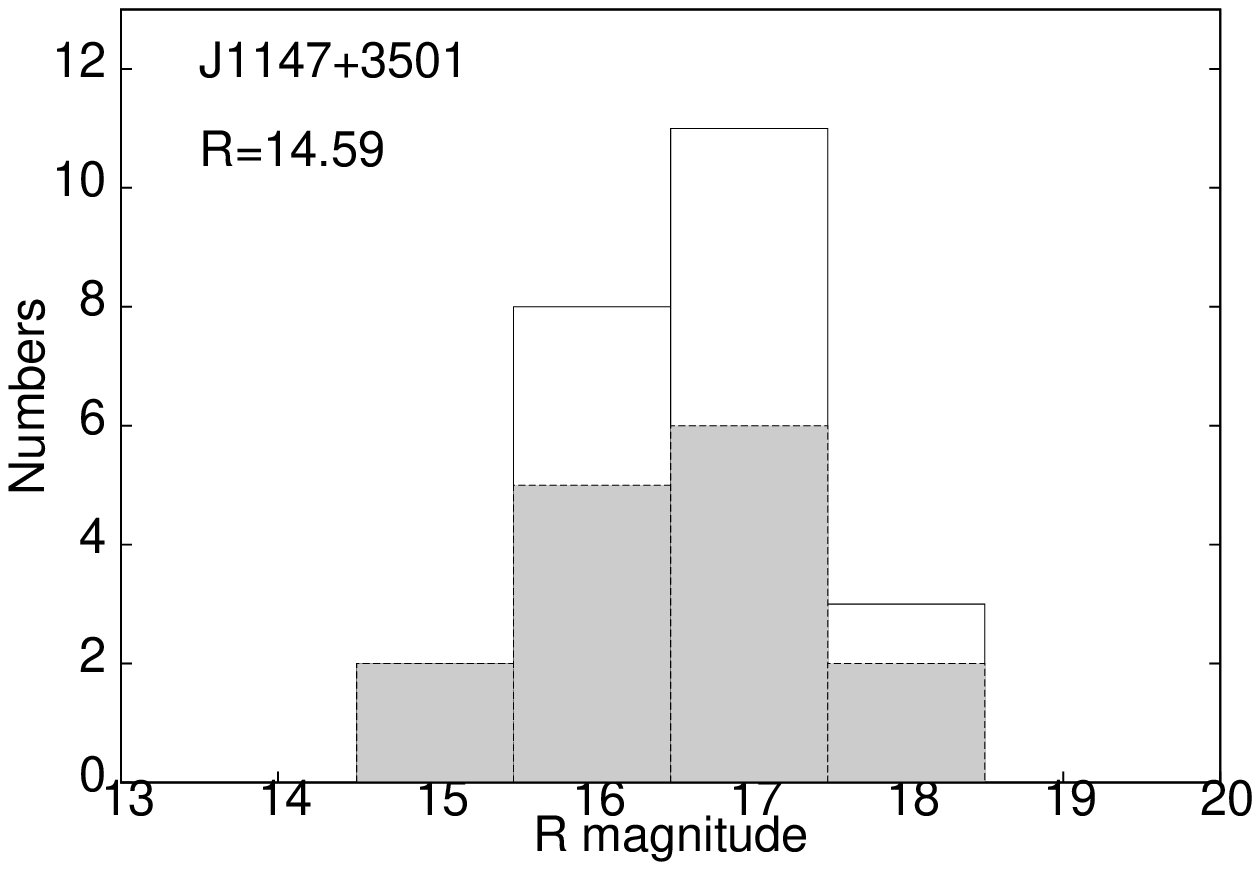,width=1.6in,angle=0}
  \psfig{file=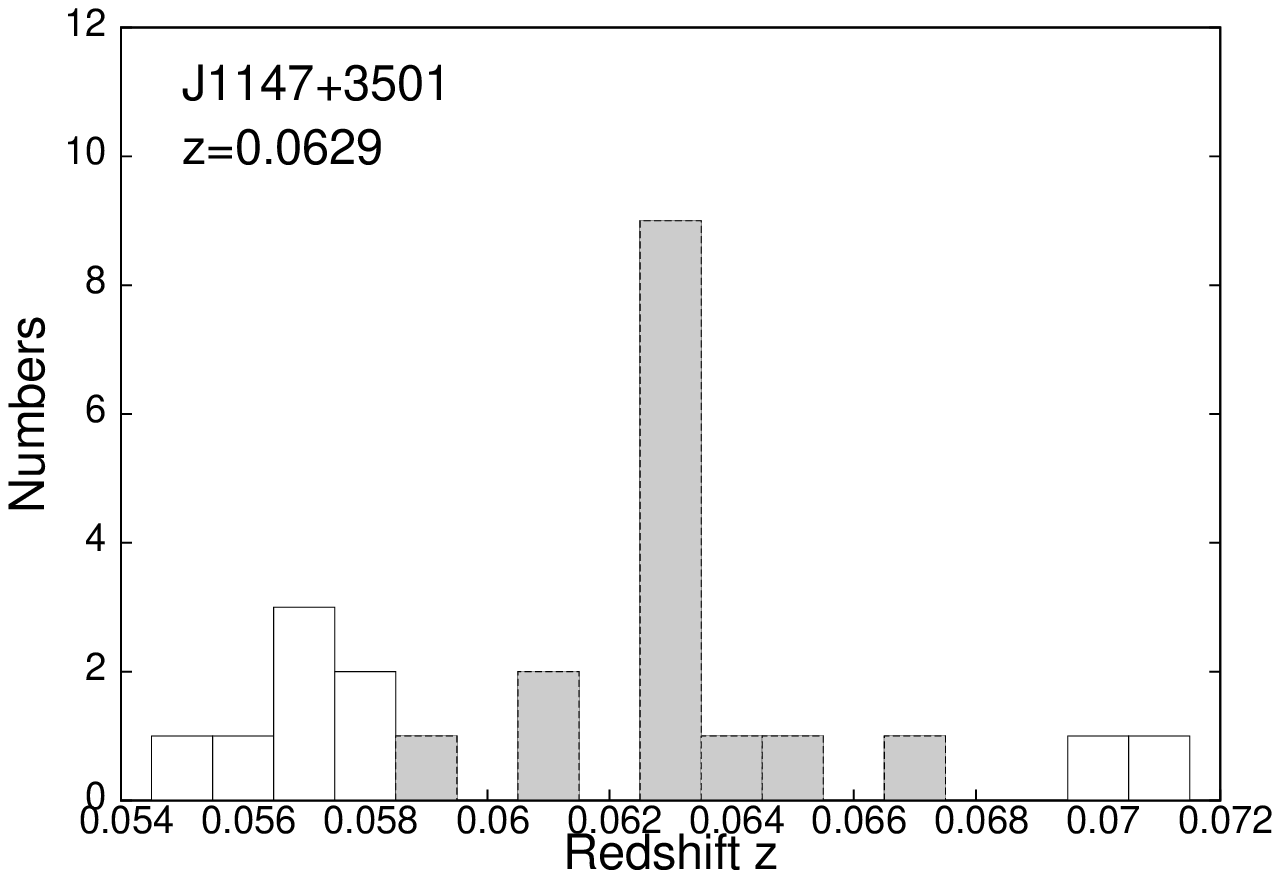,width=1.6in,angle=0}
\hspace{0.4cm}
  \psfig{file=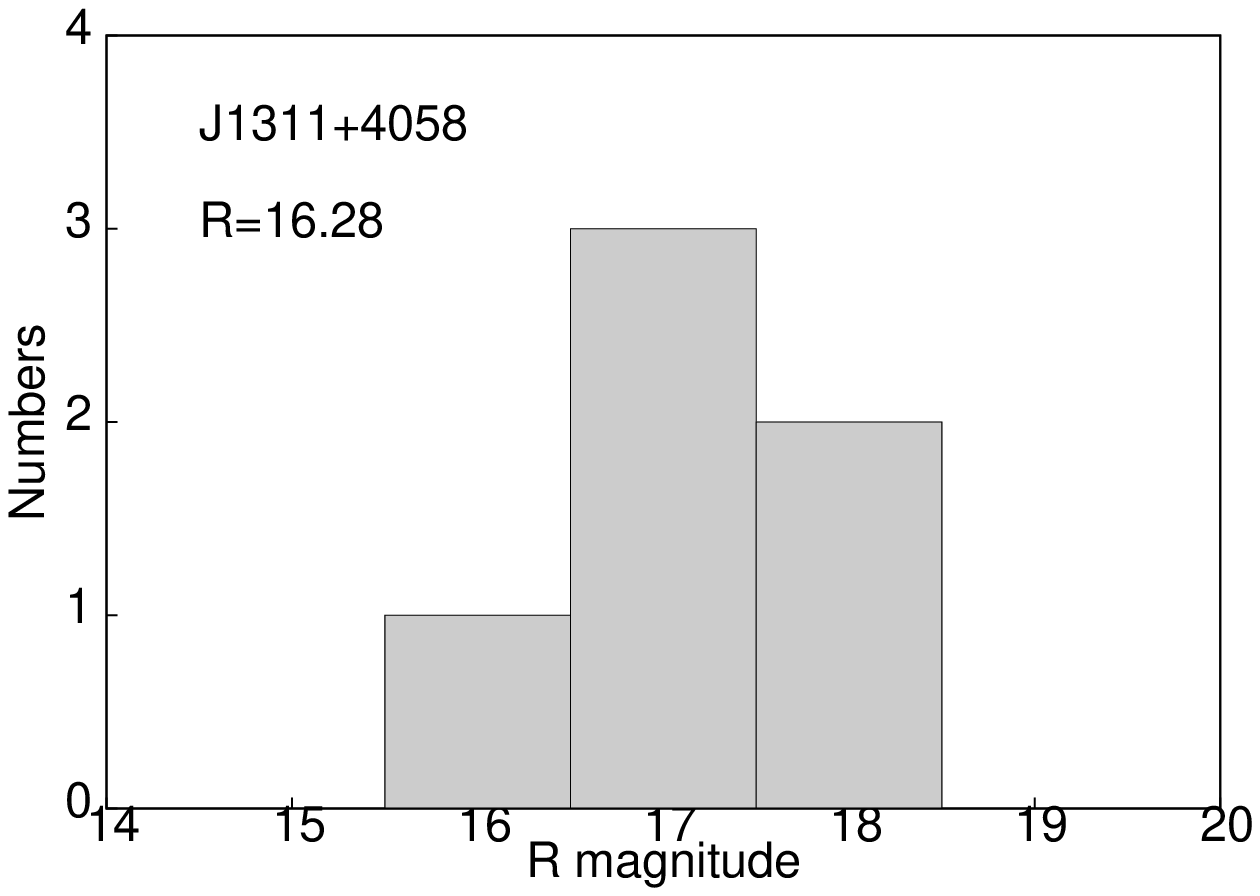,width=1.6in,angle=0}
  \psfig{file=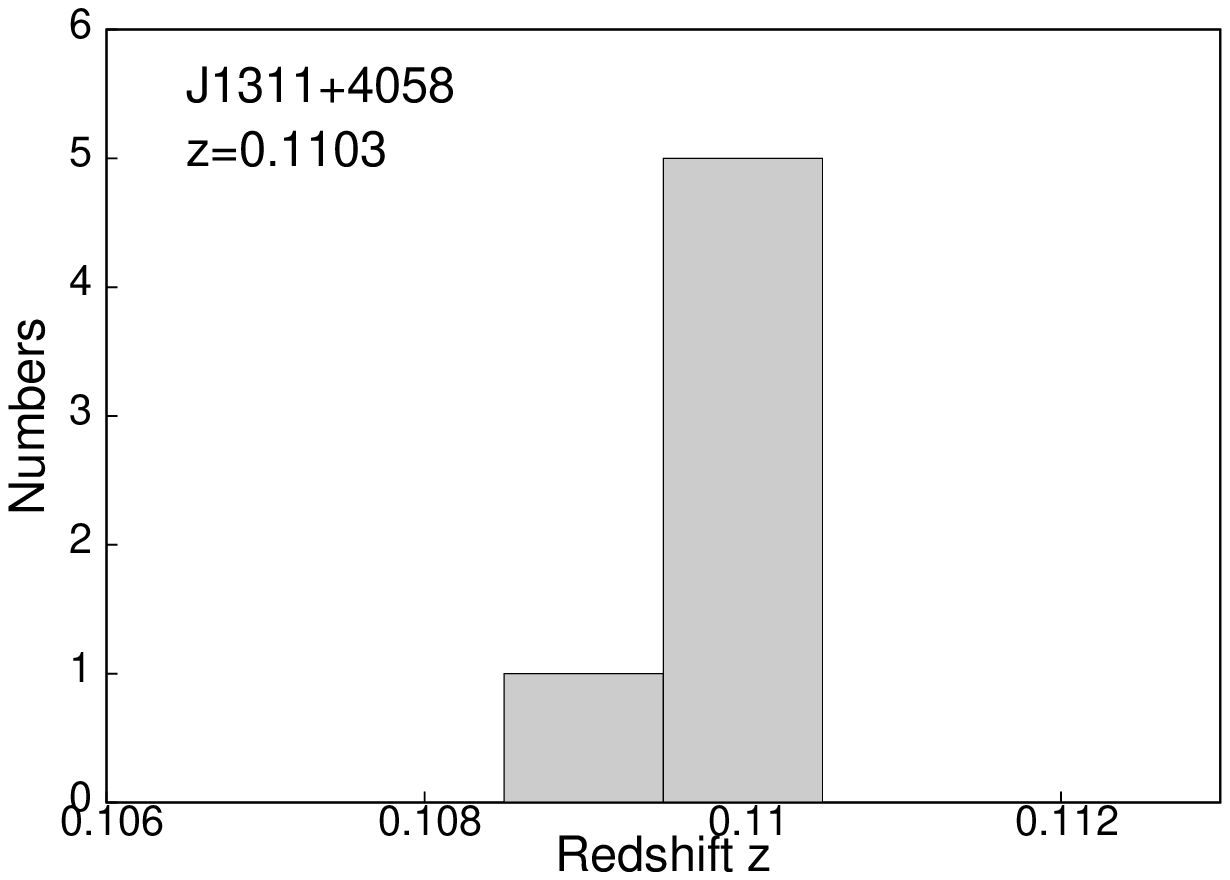,width=1.6in,angle=0}
      }
      }
\caption[]{Radio images towards J0926$+$6519, J1006$+$3454, J1147$+$3501 and J1311$+$4058.
           These are from NVSS. The symbols have the same meaning as in Fig. 1.
           }
\end{figure*}

%%%%%%%%%%%%%%%%%%%%%%%%%%%%%%%%%%%%%%%%%%%%%%%%%%%%%%%%%%%%%%%%%%%%%%%%%%%%%%%%%%%%%

%%%%%%%%%%%%%%%%%%%%%%%%%%%%%%%%%%%%%%%%%%%%%%%%%%%%%%%%%%%%%%%%%%%%%%%%%%%%%%%
\begin{figure*}
\vbox{
%\vskip -0.5cm
\hbox{
  \psfig{file=1328_0307_VLSS_3RAD_5R_2500_1500.PS,width=3.3in,angle=0}
  \psfig{file=1400_3019_3RAD_5R_2500_1500.PS,width=3.58in,angle=0}
     }
\vskip -0.5cm
\hbox{
\hspace{0.5cm}
  \psfig{file=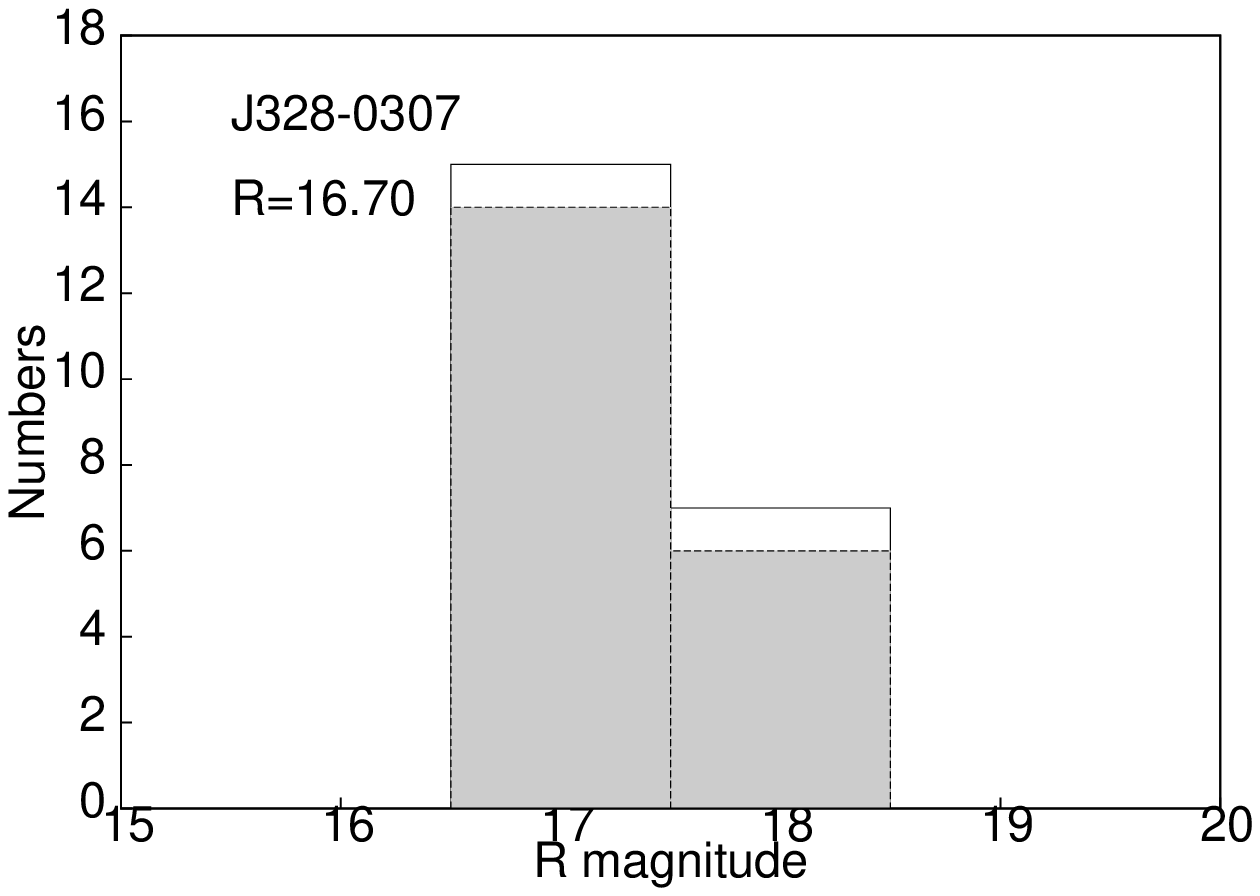,width=1.6in,angle=0}
  \psfig{file=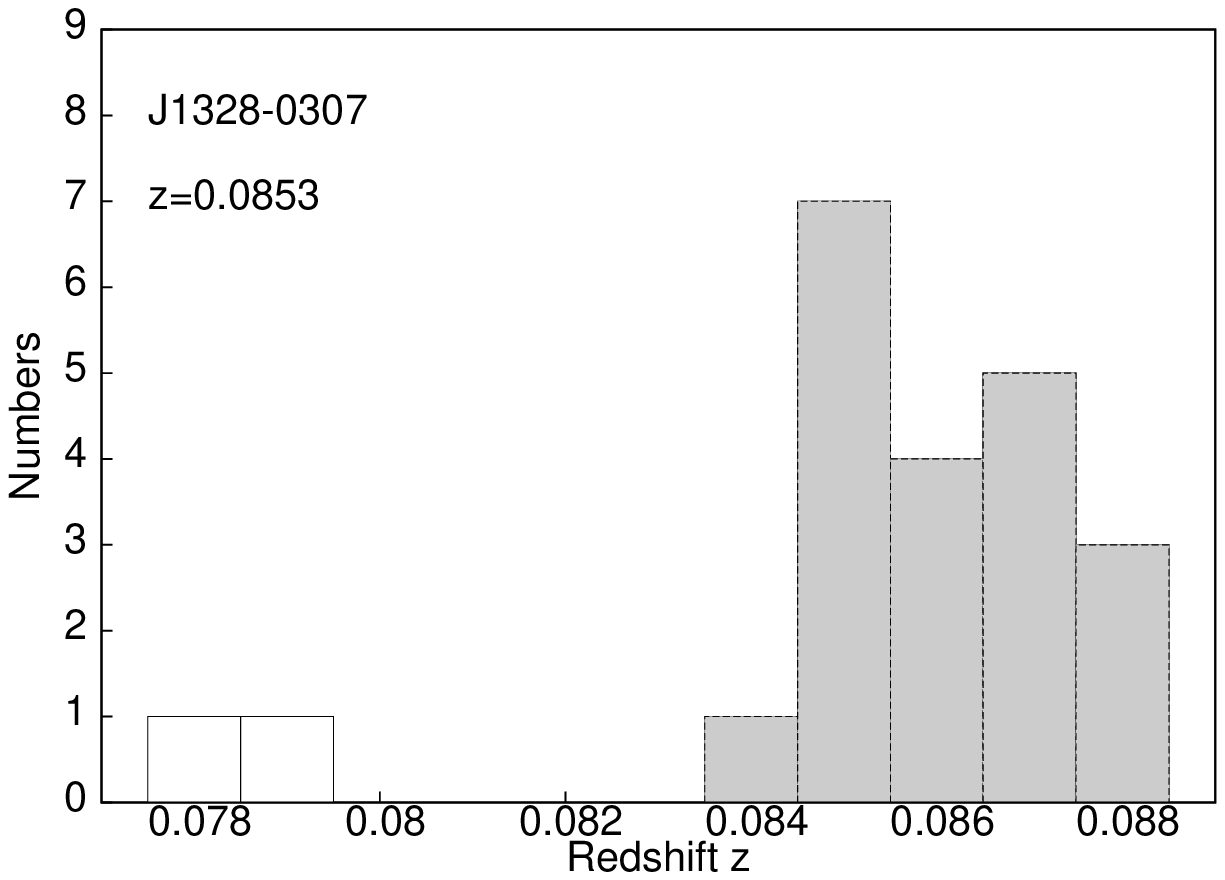,width=1.6in,angle=0}
\hspace{0.2cm}
  \psfig{file=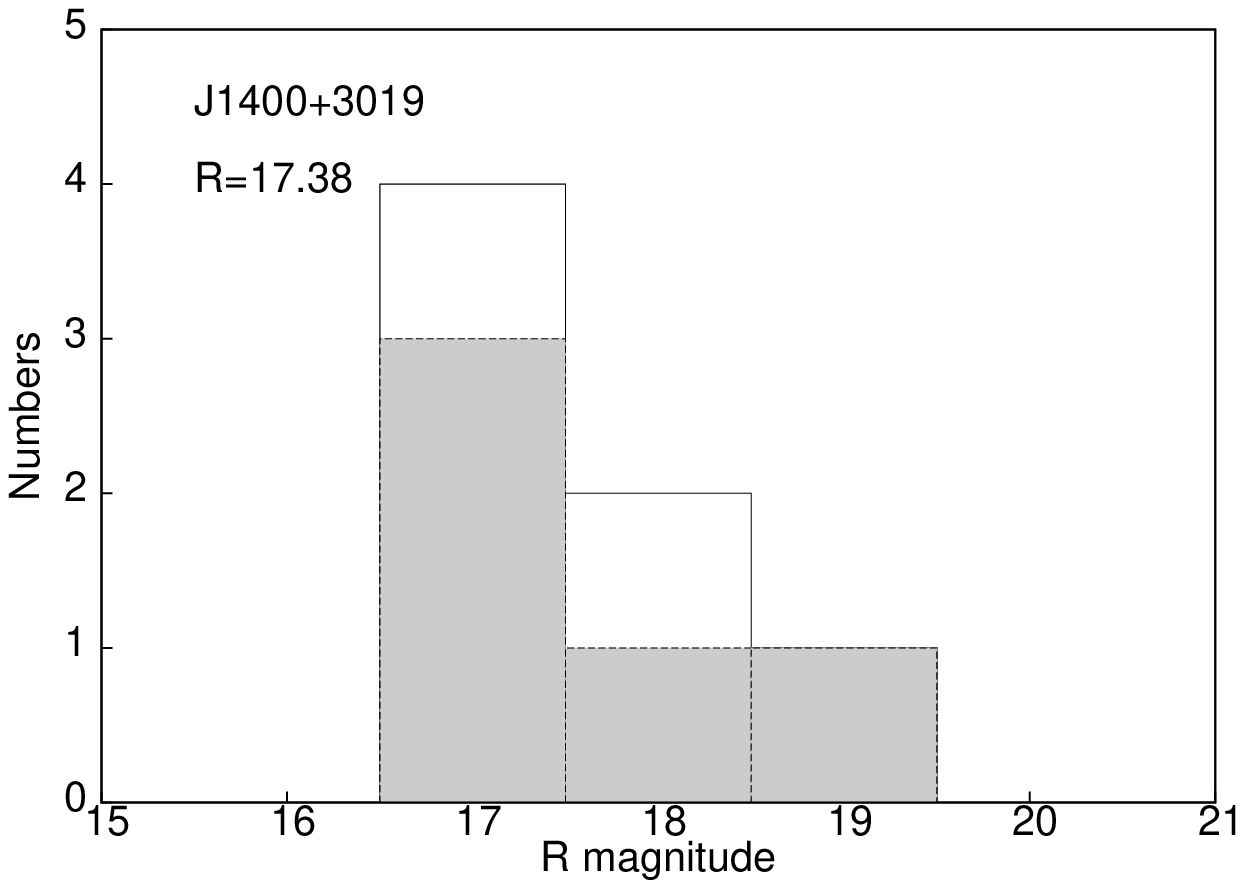,width=1.6in,angle=0}
  \psfig{file=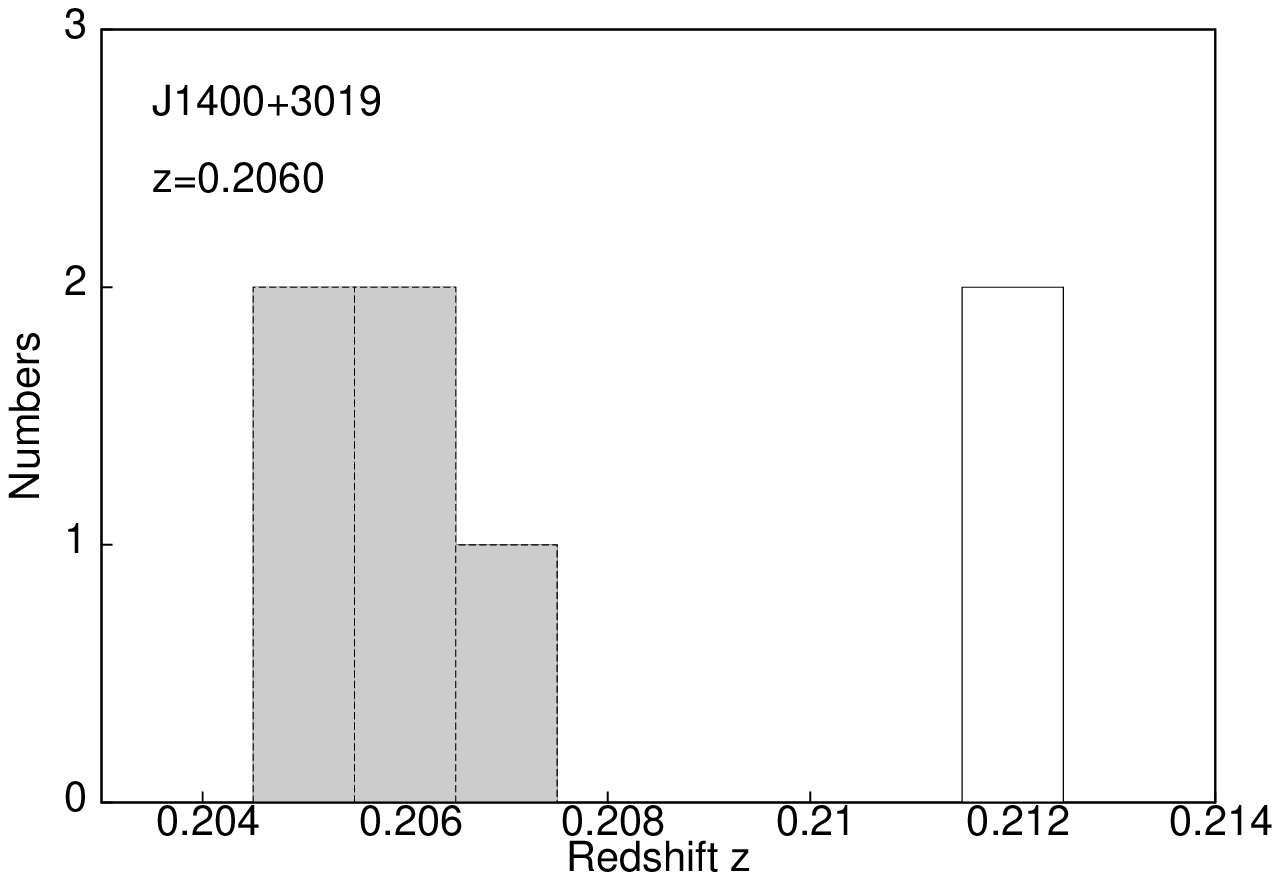,width=1.6in,angle=0}
      }
\vskip -0.1cm
\hbox{
  \psfig{file=1428_2918_3RAD_5R_2500_1500.PS,width=3.45in,angle=-90}
  \psfig{file=1635_3608_3RAD_5R_2500_1500.PS,width=3.3in,angle=-90}
     }
\vskip -0.3cm
\hbox{
\hspace{0.4cm}
  \psfig{file=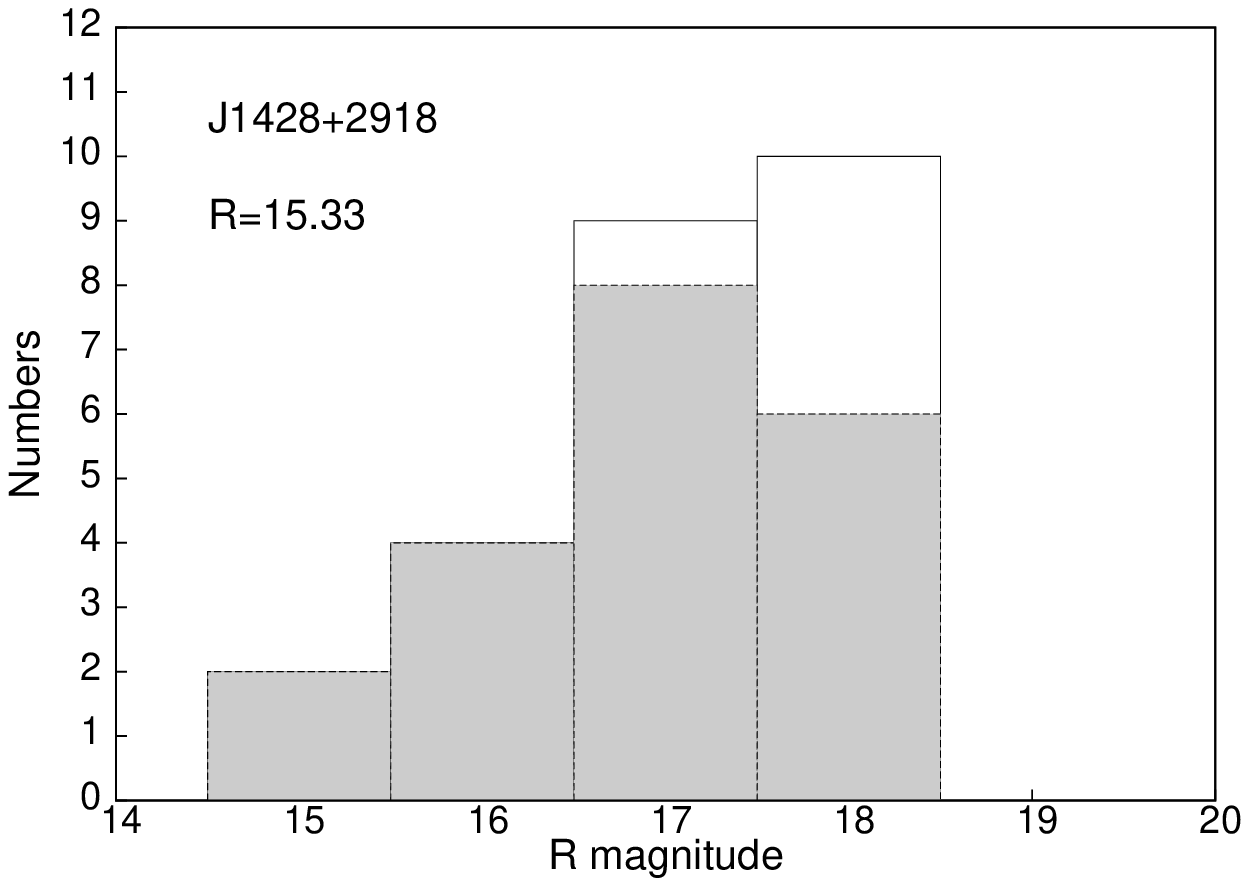,width=1.6in,angle=0}
  \psfig{file=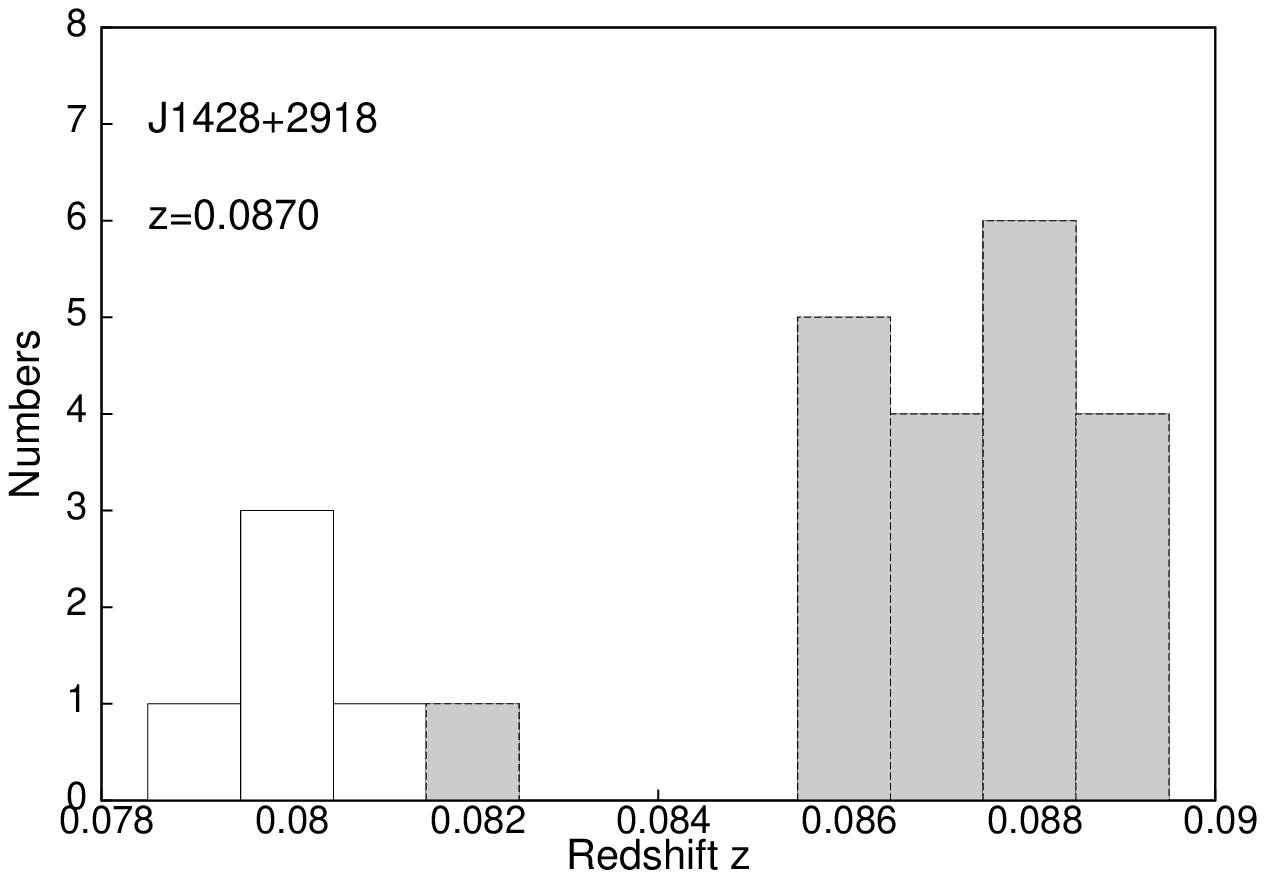,width=1.6in,angle=0}
\hspace{0.3cm}
  \psfig{file=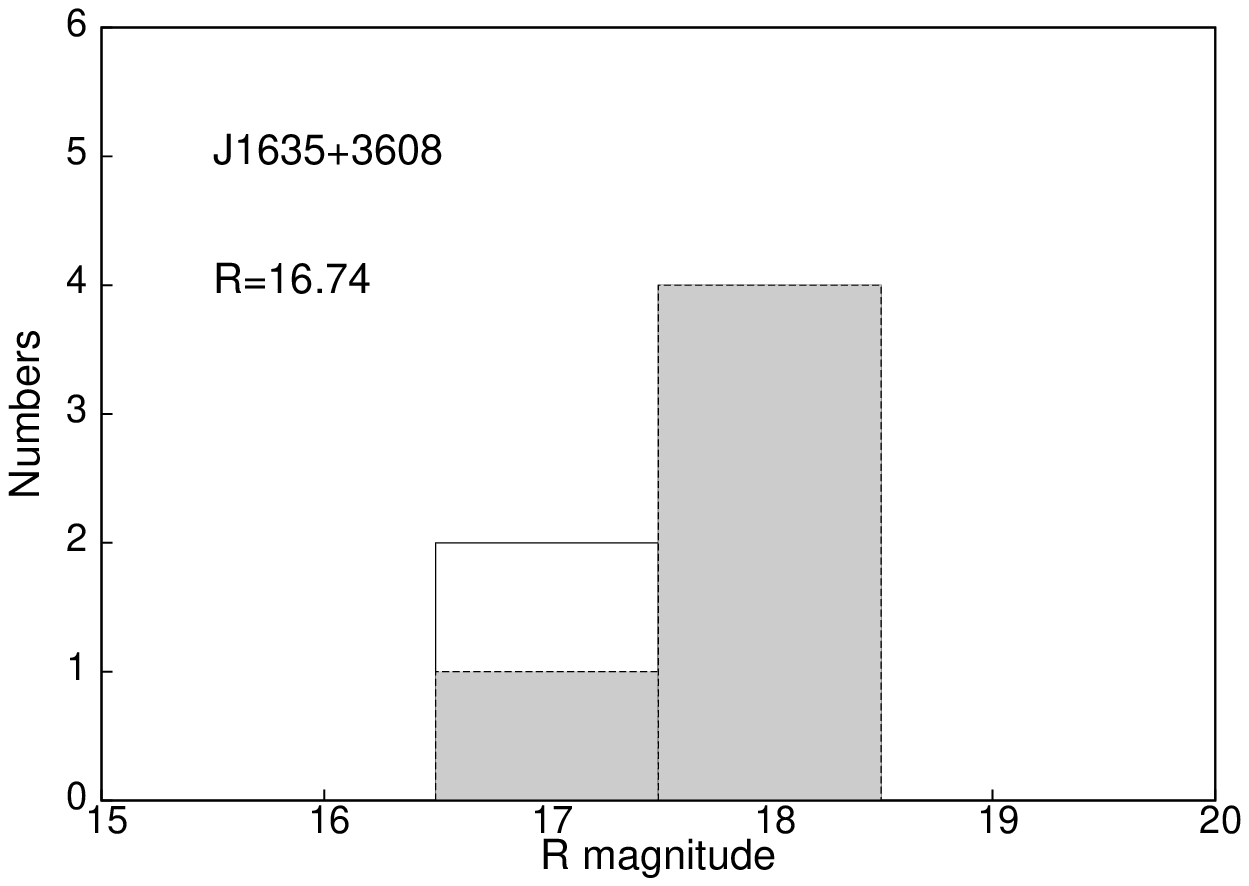,width=1.6in,angle=0}
  \psfig{file=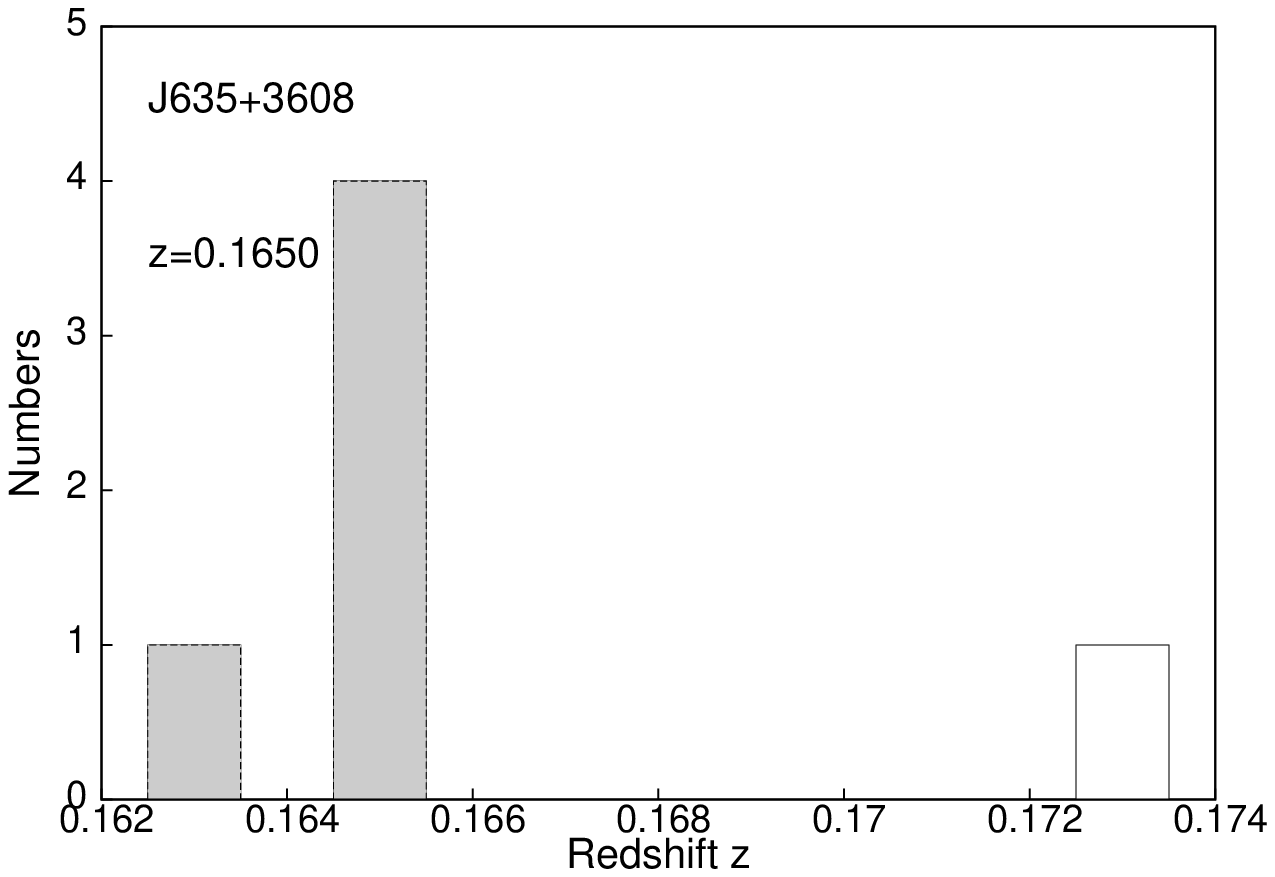,width=1.6in,angle=0}
     }
      }
\contcaption{Radio images towards J1328$-$0307, J1400$+$3019, J1428$+$2918 and J1635$+$3608. 
            For images towards J1328$-$0307 and J1400$+$3019, the grey scale represents
            the NVSS and the contours represent the VLSS images, while for 
            J1428$+$2918 and J1635$+$3608 the images are from NVSS. The symbols have 
            the same meaning as in Fig. 1.
           }
\end{figure*}

%%%%%%%%%%%%%%%%%%%%%%%%%%%%%%%%%%%%%%%%%%%%%%%%%%%%%%%%%%%%%%%%%%%%%%%%%%%%%%%%%%%%%%%%

%%%%%%%%%%%%%%%%%%%%%%%%%%%%%%%%%%%%%%%%%%%%%%%%%%%%%%%%%%%%%%%%%%%%%%%%%%%%%%%

\section{Results and Discussion}
Almost all the giant radio sources, except for
J1113+4017 which is associated with Abell 1203 of richness class 0
(Abell, Corwin \& Olowin 1989), are either in poor groups or are relatively 
isolated,  consistent with earlier studies.  We first focus 
on the two sources  J1113$+$4017 and J1552$+$2005 
which either exhibit some evidence of filamentary-like structure in the
distribution of galaxies or a higher projected density of galaxies (Fig. 1).
We consider the galaxy distributions to form a coherent filamentary-like
structure if it is at least 5 Mpc long, with the ratio of the length to width 
being $\gapp$5. Within this structure, the surface density of galaxies
should be $\gapp$2 Mpc$^{-2}$ and the surface density contrast with the 
rest of the field should be $\gapp$10.   

In the GRG J1552+2005 (3C326) the galaxies within $\pm$1500 km s$^{-1}$
form a filamentary structure. There is a concentration of galaxies on the outer side of 
the eastern lobe, which is both closer and brighter, demonstrating that 
the radio structure has been influenced by interaction with this large-scale 
structure. It has the largest arm-length ratio (R$_{\theta}$=3.10) amongst our 
sample of sources, and as expected from interaction of the radio jet with the 
cluster of galaxies, the nearer lobe is brighter (see Eilek \& Shore 1989;
Jeyakumar et al. 2005). The galaxies with relative velocities of $+$1500 km s$^{-1}$ 
to $+$2500 km s$^{-1}$ appear quite distinct in the velocity distribution of the galaxies, 
and could be part of a background structure.

Although J1113+4017 occurs in a rich cluster of galaxies, there is no concentration 
of galaxies in the immediate vicinity of the radio lobes, and the source appears 
reasonably symmetric with arm-length and flux density ratios of the lobes being $<$1.5.
We have examined the distribution of galaxies in different velocity slices to examine
possible evidence of filamentary structures, but do not find evidence of any
feature that meets our criteria. 

Fig. 2 shows the radio contour images of eight sources along with the positions
of the galaxies within $\pm$2500 km s$^{-1}$. Six radio sources which occur in 
environments of very low galaxy density with $\leq$2 optical galaxies in the field 
(see Table 2) are not shown here. While visually examining the distributions of galaxies
there appeared to be coherent structures in a few cases, such as J1006+3454, J1400+3019 
and J1635+3608, but these did not satisfy our criteria for classifying these as
filamentary-like structures. 

Of the sources in our sample, the most asymmetric source is J0313+4120, which exhibits the
highest degree of flux density asymmetry. This source is in a low-density environment with 
only one galaxy within $\pm$2500 km s$^{-1}$  and located $\sim$2.7 Mpc from the parent galaxy. 
However, it has a strong and variable radio core, and its observed asymmetry is likely to be 
significantly influenced by relativistic motion. This may also be the case for J1147+3501, 
where  most of the galaxies which are within $\pm$1500 km s$^{-1}$ of the identified GRG
lie in the vicinity of the radio source. It has been suggested by Schoenmakers et al.
(1999) that there are two radio sources in the centre of the field, with the two components on the 
north-western side forming an independent source, J1147+3503, with a linear size of 
153 kpc. The GRG is reasonably symmetric in the location of the lobes (R$_{\theta}$=1.20), 
although the flux density ratio is 2.09. A variable and prominent core (Schoenmakers
et al. 1999; see also Table 3) suggests that the flux density asymmetry could be due to 
effects of relativistic motion.  There is no concentration of galaxies in the outer 
extremities of the lobes.
 
Besides J0313+4120 and J1147+3501, the two other sources with prominent central components
are J1006+3454 and J1247+6723. However high-resolution observations show the latter two to 
be rejuvenated radio sources with compact inner doubles (see Saikia \& Jamrozy 2009, and 
references therein). In all the sources, except for J1147+3501, the nearer component tends to be brighter.  
Assuming that the oppositely-directed jets are intrinsically symmetric, this 
suggests that although these sources are often in regions of low galaxy density, 
the source asymmetries are affected by density asymmetries of the IGM on these scales.

\section{Concluding remarks}
The most significant finding of this study is that in the GRG
J1552+2005 (3C326) which has the largest arm-length ratio of 3.10, the
shorter arm is interacting with a group of galaxies which forms part
of a filamentary structure. Except for one source, J1113+4017,
which occurs in a cluster of galaxies of Abell richness class 0, most
occur in regions of low galaxy density. However, in these sources too
the shorter arm tends to be brighter suggesting that these are affected
by IGM density asymmetries which are not apparent in density counts of
galaxies.  In two cases, J0313+4120 and J1147+3501, which have
prominent and variable radio cores, the observed flux density ratios are
also likely to be influenced by the effects of relativistic motion. It would 
be useful to extend such studies to a larger sample of sources and also probe 
their environments via sensitive X-ray observations.

%%%%%%%%%%%%%%%%%%%%%%%%%%%%%%%%%%%%%%%%%%%%%%%%%%%%%%%%%%%%%
\section*{Acknowledgments}
We thank the reviewer for detailed comments and a careful reading
of the manuscript which have helped improve the paper, and Tirthankar Roy Choudhury
for useful discussions.
AP thanks NCRA, TIFR for hospitality during the course of this work.
SDSS-III is managed by the Astrophysical Research Consortium for the Participating Institutions of 
the SDSS-III Collaboration.
Funding for SDSS-III has been provided by the Alfred P. Sloan Foundation, the Participating Institutions, 
the National Science Foundation, and the U.S. Department of Energy Office of Science. 
This research has made use of the NASA/IPAC extragalactic database (NED)
which is operated by the Jet Propulsion Laboratory, Caltech, under contract
with the National Aeronautics and Space Administration. We thank numerous contributors
to the GNU/Linux group. This research has made use of NASA's Astrophysics Data System.
%%%%%%%%%%%%%%%%%%%%%%%%%%%%%%%%%%%%%%%%%%%%%%%%%%%%%%%%%%%%%%%%%%%%%%%%%%%%%%%%%%%%%%%%%%%%%%%%%%%%%%%%%%%%%%%

{}

%%%%%%%%%%%%%%%%%%%%%%%%%%%%%%%%%%%%%%%%%%%%%%%%%%%%%%%%%%%%%%%%%%%%%%%%%%%%%%

\end{document}